\documentclass[aps,showpacs,11pt]{revtex4}
\usepackage{dcolumn}
\usepackage{graphicx}
\usepackage{amsmath}
\usepackage{amsfonts}
\usepackage{amssymb}
\usepackage{psfrag}
\usepackage{wrapfig}
\usepackage{subfigure}
\usepackage{makeidx}
\usepackage{bm}
\usepackage{epsf}
\usepackage{hyperref}
\usepackage{color}
\usepackage{multirow}

\begin{document}
	
	\title{Spherical black holes with minimally coupled scalar cloud/hair in Einstein-Born-Infeld gravity}
	
	\author{Shao-Jun Zhang}
	\email{sjzhang@zjut.edu.cn}
	\affiliation{Institute for Theoretical Physics $\&$ Cosmology, Zhejiang University of Technology, Hangzhou 310032, China \\
	United Center for Gravitational Wave Physics, Zhejiang University of Technology, Hangzhou 310032, China}
	\date{\today}

	\begin{abstract}
		\indent Previous studies showed that, in the presence of a simple and well-motivated self-interaction scalar potential, asymptotically flat and spherical black holes can carry minimally coupled and charged scalar cloud/hair in Einstein-Maxwell gravity. We extend these studies to Einstein-Born-Infeld gravity to consider the effect of nonlinearity of the electromagnetic field. Series of spherical cloudy/hairy black hole solutions are constructed numerically. Results show that increasing the Born-Infeld coupling constant $b$ will make the domain of existence of the solution shrink or even disappear when $b$ is large enough. This implies that, competing with the gravitation, nonlinearity of the electromagnetic field will make the formation of scalar cloud/hair harder or even impossible.
		
	\end{abstract}
	

	\maketitle
	
	\section{Introduction}
	
	With the advent of astrophysical observations in recent years, especially the first-ever detection of gravitational waves \cite{Abbott:2016blz,Abbott:2016nmj,Abbott:2017gyy} and photoing of black hole shadow \cite{EventHorizonTelescope:2019dse}, black hole (BH) physics has entered a golden era. It enables us to test gravity theories and probe fundamental fields in strong gravity regime with unprecedented precision \cite{Barack:2018yly}. Central to the understanding of BH nature relies on a number of no-hair theorems (for a review  see \cite{Bekenstein:1998aw,Robinson:2004zz,Chrusciel:2012jk}) which claim that asymptotically flat and stationary BH solutions in Einstein-Maxwell gravity are characterized by only three physical parameters: their mass, charge and angular momentum, and it is impossible for these BHs to carry scalar or other types of hair. However, the establishment of these theorems requires rather strong preconditions, so it is possible to circumvent them by violating the preconditions and thus BH solutions with hair can be constructed. Amounts of effort have been devoted on this direction in past decades and lots of hairy BH solutions have been found (for a review see \cite{Herdeiro:2015waa}). Here we would like to mention some of them which are most relevant to our present work. In refs. \cite{Hod:2012px,Herdeiro:2014goa}, working in the framework of Einstein's gravity, the authors found surprisingly that Kerr BH can be endowed with a massive, regular, and minimally coupled scalar cloud. When back-reacting on the geometry, the scalar cloud becomes hair and hairy rotating BHs are formed. The metric remains stationary and axially symmetric while the complex scalar field takes a harmonic time-dependence form $\Psi \sim \exp(i \omega t)$ thus circumventing the no-hair theorems. Existence of the cloud/hair requires the so-called synchronization condition, i.e., $\omega = \omega_c$. Noting that $\omega_c$ is just the marginal frequency for the rotational superradiance of Kerr BH under scalar perturbation, the scalar cloud is thus interpreted to be formed through the superradiance and the final hairy BHs should be the end state of the rotational superradiant instability \cite{East:2017ovw,Herdeiro:2017phl} (also see a review \cite{Brito:2015oca} on this topic). Moreover, this kind of scalar hair is primary and thus can be arbitrary small, so the solution reduces to the bald Kerr BH with cloud/hair decreasing. Later, the work is extended on several directions, such as considering self-interactions of the scalar field \cite{Herdeiro:2015tia,Delgado:2020hwr}, rotating and charged situation \cite{Delgado:2016jxq}, higher-dimensional cases \cite{Brihaye:2014nba,Herdeiro:2015kha,Herdeiro:2017oyt}, Proca hair \cite{Herdeiro:2016tmi,Santos:2020pmh} and boson stars cases \cite{Guerra:2019srj,Delgado:2020udb} etc.
	
    Recently, the above work is extended to non-rotating case. In Refs. \cite{Hong:2019mcj,Herdeiro:2020xmb,Hong:2020miv}, working in the framework of Einstein-Maxwell gravity, the authors found that spherical BHs can also be endowed with minimally coupled scalar cloud/hair. Also, the metric is static and spherically symmetric while the scalar field once again takes a harmonic time-dependence form with $\omega = \omega_c$. This time $\omega_c$ is the marginal frequency for the charged superradiance of Reissner-Nordstr\"{o}m (RN) BH under charged scalar perturbation \cite{Bekenstein:1973mi}. However, as has been well known, although there is charged superradiance, no bound states can be formed to trigger superradiant instability \cite{Hod:2012wmy,Hod:2013nn,Brito:2015oca}. So this kind of hairy BHs should not be interpreted as end state of superadiant instability of RN BHs. The scalar field considered takes a self-interaction potential which is found to be crucial for the cloud/hair emergence. Moreover, different from its rotating cousin, this kind of cloud/hair is secondary and cannot be arbitrary small: there is a mass gap, and thus the hairy solutions do not bifurcate from the bald RN BHs. The scalar potential considered in these work takes a polynomial form. For other types of potentials, such kind of hairy BH is also found to exist \cite{Mai:2020sac,Brihaye:2021phs}. See also refs. \cite{Brihaye:2020vce,Garcia:2021pzd,Brihaye:2021mqk} for more work on this topic .
    
    Based on these work, it is natural to extend the above studies to Einstein-Born-Infeld gravity to include the effect of non-linearity of electromagnetic field. As a modification of the usual Maxwell electrodynamics, Born-Infeld term is proposed originally to solve the divergence problem of the self-energy of point-like charges \cite{Born:1933pep,Born:1934gh} (see also the review \cite{BeltranJimenez:2017doy}).  It is then found to arise naturally from other theoretically appealing theories, like e.g. string theory which is considered to be the most promising candidate of quantum gravity theories \cite{Polchinski:1998rq,Polchinski:1998rr}. We will construct spherical BH solutions with scalar cloud/hair within this framework and focus on  studying the effects of Born-Infeld coupling constant on the formation of the scalar cloud/hair.

    The work is organized as follows. In Sec.II, we will give a brief introduction of the Einstein-Born-Infeld-scalar model and equations of motions are derived therein. Also, asymptotic behaviors of the solutions and symmetries of the model are discussed for later use. In Sec.III, series of spherical BH solutions with scalar cloud/hair are constructed numerically and their properties are analyzed. The last section is the Summary and Discussions.

	\section{The Einstein-Born-Infeld-scalar Model}
	
	In this work, we consider the Einstein-Born-Infeld theory with a minimally coupled complex scalar field. The action is
	\begin{eqnarray}\label{action}
	S = \int d^4 x \sqrt{-g} \left[\frac{R}{16\pi G} +L_{\rm BI}-(D_\mu \Psi)^\ast D^\mu \Psi -U(|\Psi|)\right],
	\end{eqnarray}
	where $G$ is the gravitational constant, R is the Ricci scalar, $\Psi$ is a complex scalar field with self-interaction potential $U(|\Psi|)$ and `$\ast$' denotes complex conjugate. The Lagrangian of the Born-Infeld electrodynamics is \cite{Born:1933pep,Born:1934gh,BeltranJimenez:2017doy,Rasheed:1997ns,Fernando:2003tz,Dey:2004yt,Cai:2004eh,Miskovic:2008ck}
	\begin{eqnarray}
	L_{\rm BI} = \frac{1}{b} \left(1-\sqrt{1+\frac{b F^2}{2}}\right),
	\end{eqnarray}
	where the gauge field strength is defined as $F_{\mu\nu} \equiv \partial_\mu A_\nu - \partial_\nu A_\mu$ with $A_\mu$ being the gauge potential, and $F^2 \equiv F_{\mu\nu} F^{\mu\nu}$. The strength of the nonlinearity of the gauge field is characterized by the Born-Infeld coupling constant $b$. In the limit $b\rightarrow 0$, $L_{\rm BI}$ reduces to the standard Maxwell term, $L_{\rm M}= -\frac{1}{4} F_{\mu\nu} F^{\mu\nu}$. The gauge covariant derivative of the complex scalar field is defined as $D_\mu \Psi \equiv \nabla_\mu \Psi + i q A_\alpha \Psi$, where $\nabla$ is the geometric covariant derivative and $q$ is the gauge coupling constant. 
	
	As shown in Refs. \cite{Herdeiro:2020xmb,Hong:2020miv}, minimally coupled scalar-hair can exist only for the presence of the self-interaction. Within the Einstein-Maxwell-scalar model, several types of the self-interaction potentials have been considered in previous work \cite{Hong:2019mcj,Herdeiro:2020xmb,Hong:2020miv,Brihaye:2020vce,Mai:2020sac,Garcia:2021pzd,Brihaye:2021phs}, including polynomial, logarithmic and exponential types. In this work, as Refs. \cite{Hong:2019mcj,Herdeiro:2020xmb,Hong:2020miv,Brihaye:2020vce,Garcia:2021pzd,Brihaye:2021phs}, we consider the potential to take a simple polynomial type
	\begin{eqnarray}
	\quad U(|\Psi|) = \mu^2 |\Psi|^2 - \lambda |\Psi|^4 + \nu |\Psi|^6,
	\end{eqnarray}
	where $\mu, \lambda, \nu$ are free parameters with $\mu$ representing the rest mass of the scalar field. This type of potential is motivated by the studies of scalar field in flat spacetime \cite{Coleman:1985ki} and usually dubbed as Q-ball potential.
	
	The field equations can be obtained by varying the action (\ref{action})
    \begin{eqnarray}
    && R_{\mu\nu} - \frac{1}{2} g_{\mu\nu} R = 8\pi G \left[T_{\mu\nu}^{({\rm EM})} + T_{\mu\nu}^{({\rm \Psi})} \right],\label{EinsteinEq}\\
    && D_\mu D^\mu \Psi = \frac{d U}{d|\Psi|^2} \Psi,\label{ScalarEq}\\
    && \nabla_\mu \left(\frac{F^{\mu\nu}}{\sqrt{1+\frac{b F^2}{2}}}\right) = i q \left[\Psi (D^\nu \Psi)^\ast- \Psi^\ast D^\nu \Psi \right],\label{BornInfeldEq}
    \end{eqnarray}
    where the two components of the energy-momentum tensor are
    \begin{eqnarray}
    &&T_{\mu\nu}^{({\rm EM})} = \frac{1}{\sqrt{1+\frac{b F^2}{2}}} F_{\mu}^{~\rho} F_{\nu \rho} + \frac{1}{b} g_{\mu\nu} \left(1-\sqrt{1+\frac{b F^2}{2}}\right),\\
    &&T_{\mu\nu}^{({\rm \Psi})} = D_\mu \Psi (D_\nu \Psi)^\ast + (D_\mu \Psi)^\ast D_\nu \Psi - g_{\mu\nu} \left[D_\rho \Psi (D^\rho \Psi)^\ast + U(|\Psi|)\right].
    \end{eqnarray}

    To search spherical charged BHs with scalar hair, we consider the following ansatz
    \begin{eqnarray}
    &&ds^2 = -N(r) \sigma(r)^2 dt^2 + \frac{dr^2}{N(r)} + r^2 (d\theta^2 + \sin^2 \theta d\varphi^2),\\
    && A_\mu=(V(r), 0, 0, 0), \quad \Psi =\psi(r) e^{- i \omega t},
    \end{eqnarray}
    where $\omega$ is the real oscillation frequency of the scalar field. Inserting this ansatz into the field equations (\ref{EinsteinEq}) (\ref{ScalarEq}) (\ref{BornInfeldEq}), we have 
    \begin{eqnarray}\label{EOMs}
     &&N' = - 8\pi G r \left[\frac{1}{b} \left(\frac{1}{\sqrt{1-\frac{b V'^2}{\sigma^2}}} - 1\right) + N \psi'^2 + U(\psi) + \frac{(\omega - q V)^2 \psi^2}{N \sigma^2} \right] + \frac{1-N}{r},\nonumber\\
     &&\sigma' = 8\pi G r \sigma \left[ \psi'^2 + \frac{(\omega - q V)^2 \psi^2}{N^2 \sigma^2} \right],\nonumber\\
     &&V'' =  \left( \frac{\sigma'}{\sigma} - \frac{2}{r} + \frac{2 b  V'^2}{r\sigma^2}\right) V' - \frac{2 q (\omega - q V) \psi^2 }{N} \left(1- \frac{b V'^2}{\sigma^2}\right)^{3/2},\nonumber\\
     &&\psi'' = -\left(\frac{2}{r}+\frac{N'}{N} + \frac{\sigma'}{\sigma}\right) \psi' - \frac{(\omega -q V)^2 \psi }{N^2 \sigma^2} + \frac{1}{2 N} \frac{dU}{d\psi},
    \end{eqnarray}
    where the prime denotes derivative with respect to the radial coordinate $r$.
    
    Before diving into the numerical integration of the field equations to construct spherical charged BHs with scalar hair, for later use, let us first obtain the asymptotic solution of the four radial functions, $\{ N(r), \sigma(r), V(r), \psi(r) \}$, both at the horizon and spatial infinity. Also, we will fix the symmetries of the field equations to simplify our calculations.
    
	\subsection{Asymptotic solution}
	
	In this work, we are interested in searching non-extremal spherical charged BHs with regular scalar hair. So, near the horizon $r=r_h$ defined as $N(r_h)=0$, the four radial functions are assumed to have Taylor expansions as
	\begin{eqnarray}\label{NearHorzion}
	&&N(r) = N_1 (r - r_h) + \cdots,\quad \sigma(r)=\sigma_h + \sigma_1 (r-r_h) + \cdots,\nonumber\\
	&&V(r) = V_h + V_1 (r-r_h) + \cdots,\quad \psi(r) = \psi_h + \psi_1 (r-r_h) + \cdots,
	\end{eqnarray}
	where `$\cdots$' denotes higher-order terms. The field equations should be satisfied order by order near the horizon, which gives us the following relations between the expansion coefficients
	\begin{eqnarray}
	&&N_1= \frac{1}{r_h} + 8\pi G r_h \left[\frac{1}{b} \left(1 - \frac{1}{\sqrt{1- \frac{b V_1^2}{\sigma_h^2}}}\right) - U(\psi_h)\right], \\
	&&\sigma_1= \frac{8 \pi G r_h \sigma_h}{N_1^2} \left(\frac{q^2 V_1^2 \psi_h^2}{\sigma_h^2} + \frac{U'(\psi_h)^2}{4}\right), \nonumber\\
	&& \psi_1 = \frac{U'(\psi_h)}{2 N_1},
	\end{eqnarray}
	and also the so-called synchronization condition
	\begin{eqnarray}\label{SynchronizationCondition}
	\omega = q V_h, 
	\end{eqnarray}
	which is the same as in the Maxwell case considered in Ref. \cite{Herdeiro:2020xmb,Hong:2020miv}. It should be noted that this is just the threshold condition for the occurrence of superradiance of Born-Infeld BH under charged scalar perturbations \cite{Rahmani:2020lju}. 
	So up to first order expansions in (\ref{NearHorzion}), there are left only four free parameters,  $\{\sigma_h, V_h, V_1, \psi_h\}$, characterizing the near-horizon behaviors of the solution. However, they are still not totally independent as we will show later. Once their values are given, the near-horizon behaviors and then the full solution (if exists) are determined uniquely.
	
    With a similar analysis, the asymptotic behaviors of the four radial functions at spatial infinity $r \rightarrow \infty$ can also be obtained
	\begin{eqnarray}\label{SpatialInfinity}
	&&N(r) = 1 - \frac{2 G M}{r} + \frac{4 \pi G Q_e^2}{r^2}\cdots, \nonumber\\
	&&\sigma(r) =1- \frac{4 \pi G c_0^2 \mu^2}{\mu_\infty r} e^{-2\mu_\infty r} + \cdots,\nonumber\\
	&&V(r) = \Phi - \frac{Q_e}{r} + \cdots,\nonumber\\
	&&\psi(r) = c_0 \frac{e^{-\mu_\infty r}}{r} + \cdots,
	\end{eqnarray}
	where $\mu_\infty \equiv \sqrt{\mu^2 - (\omega - q \Phi)^2}$. 
	Also, there are four free parameters $\{M, Q_e, \Phi, c_0\}$, which are the ADM mass, the total electric charge, the chemical potential and an arbitrary constant respectively, characterizing the asymptotic behaviors of the solution at infinity. As we are searching asymptotically flat solutions, at infinity the metric should approach Minkowski metric while both the gauge field strength and the scalar field vanish, so we should have the following bound state condition 
	\begin{eqnarray} \label{BoundStateCondition}
	|\omega - q\Phi| \leq \mu.
	\end{eqnarray}

	\subsection{Symmetries fixing}
	
	The model possess several symmetries. First, it is obvious that the model is invariant under the gauge transformation 
	\begin{eqnarray}\label{Gauge}
	\omega \rightarrow \omega + \xi,\quad V \rightarrow V + \frac{\xi}{q}.
	\end{eqnarray}
	With this gauge freedom, it is convenient to choose the gauge condition
	\begin{eqnarray}\label{GaugeCondition}
	\omega = 0,
	\end{eqnarray}
	which, combing with the synchronization condition (\ref{SynchronizationCondition}), implies the vanishing of the gauge field at the horizon $V_h = 0$. After taking this gauge condition, there remain three scaling symmetries of the field equations:
	\begin{eqnarray}
	&(i)  &t \rightarrow a t, \quad V \rightarrow V/a,\quad \sigma \rightarrow \sigma/a,\\
	&(ii) &r \rightarrow a r, \quad q \rightarrow q/a, \quad \mu \rightarrow \mu/a, \quad \lambda \rightarrow \lambda/a^2, \quad \nu \rightarrow \nu/a^2, \quad b \rightarrow a^2 b,\\
	&(iii) &V \rightarrow a V, \quad \psi \rightarrow a \psi, \quad q \rightarrow q/a, \quad \lambda \rightarrow \lambda/a^2, \quad \nu \rightarrow \nu/a^4, \quad G \rightarrow G/a^2, \quad b \rightarrow b/a^2,
	\end{eqnarray}
	where the non-zero scaling parameter $a$ is arbitrary. The first symmetry can be fixed by imposing the boundary condition $\sigma(\infty)=1$. Practically, to make numerical calculations simply, we will relax the boundary value $\sigma(\infty)$ and set $\sigma_h =1$ at first to obtain numerical solution, and then apply the first scaling to transform the solution to satisfy $\sigma(\infty)=1$. Due to the second and the third symmetries, the model only depends on four dimensionless input parameters $\frac{\nu \mu^2}{\lambda^2}, \frac{b \mu^4}{\lambda}, \frac{q}{\sqrt{\lambda}}$ and 
	\begin{eqnarray}
	\alpha \equiv \frac{4 \pi  G \mu^2}{\lambda}.
	\end{eqnarray}
	In practical calculations, equivalently, we would like to use these two scaling symmetries to fix the parameters $\mu=1$ and $\lambda=1$.
	
   Also, without loss of generality, as Ref. \cite{Brihaye:2020vce}, we would like to fix $\nu = \frac{9}{32}$ and $r_h=0.15$. 
    
   \section{Numerical results}
   
   From the above discussions, we know that after fixing the symmetries and choosing $\nu = \frac{9}{32}$ and $r_h=0.15$, there are left only two near-horizon parameters $\{V_1, \psi_h\}$ beyond the three coupling constants $\{\alpha, b, q\}$. Once their values are given, the near-horizon expansions (\ref{NearHorzion}) are fixed, and then the solution, if exists, can be obtained by solving the field equations (\ref{EOMs}) numerically.  Practically, we would like to fix $V_1$ first and evaluate the boundary conditions at $r=r_h (1+\epsilon)$ with the near-horizon expansions (\ref{NearHorzion}), where $\epsilon$ is a small deviation parameter typically taking value of order $10^{-8}$. Then we integrate the field equations (\ref{EOMs}) numerically towards a large enough ultraviolet cutoff $r_\infty$ which typically takes value of order $10^2 r_h$. However, not any $\psi_h$ maps to a solution which is asymptotically flat. We should adjust its value so that the solution satisfies the asymptotic behavior at infinity (\ref{SpatialInfinity}), which can be achieved by applying the shooting method.
   	
   	From the numerical solutions, one can read off the ADM mass, the total electric charge and the chemical potential as
   	\begin{eqnarray}
   		M = -\frac{r}{2 G} (N(r)-1)|_{r\rightarrow r_\infty},\qquad Q_e = r^2 V'(r)|_{r\rightarrow r_\infty},\qquad \Phi = V(r)|_{r\rightarrow r_\infty}.
   	\end{eqnarray}
   	The total electric charge can be decomposed into two parts $Q_e = Q_H +  Q_S$, with $Q_H = \frac{1}{4\pi} \oint_H dS_r F^{t r}$ being the electric charge within the horizon and $Q_S$ is the Noether charge carried by the scalar hair \cite{Herdeiro:2020xmb}
   	\begin{eqnarray}
   		Q_S = q \int_{r_h}^{\infty} dr \frac{2 r^2 (q V - \omega) \psi^2}{N \sigma}.
   	\end{eqnarray}
   	So a natural measure of hairiness can be defined as
   	\begin{eqnarray}
   	h \equiv \frac{Q_s}{Q_e},
   	\end{eqnarray}
   	which is $0$ for bald RN BH while being $1$ for a soliton.
   	
   	We will consider three cases with different coupling limit: (a) Fixed Born-Infeld BHs endowed with charged scalar clouds, in which the scalar field is decoupled from both the metric and electromagnetic field; (b) Fixed Schwarzschild BHs with charged scalar clouds, in which the scalar field is coupled to the electromagnetic field but both are decoupled from the metric; (c) Static spherical charged BHs with charged scalar hair, in which all three fields are coupled. In all three cases, our main focus is on the effects of the Born-Infeld coupling constant $b$ on the scalar clouds/hair. As has been shown, the parameter space is rather huge, so we should not pursue to scan the whole parameter space to give the complete existence domain of cloudy/hairy solutions. Rather, we would like to present some sequences of typical solutions which we hope can illustrate the general properties.
  
   \subsection{Charged scalar clouds on Born-Infeld BHs}
   
   Now let us first consider the limit where the scalar field is decoupled from both the metric and electromagnetic fields. This limit can be achieved by a rescaling $V \rightarrow V/\alpha, q \rightarrow \alpha q$ and letting $\alpha \rightarrow 0$. In this limit, we are left a test scalar field on a fixed background. We consider the background to be a static BH which has already been well known as the Born-Infeld BH \cite{Rasheed:1997ns,Fernando:2003tz,Dey:2004yt,Cai:2004eh,Miskovic:2008ck}
   \begin{eqnarray}\label{BornInfeldBH}
   && N(r) = 1 - \frac{2 m}{r} + \frac{2 r^2}{3 b} \left(1 - \sqrt{1 + \frac{b^2 Q^2}{r^4}}\right) + \frac{4 Q^2}{3 r^2}\ _2F_1 \left[\frac{1}{4}, \frac{1}{2}, \frac{5}{4}, -\frac{b Q^2}{r^4}\right],\qquad \sigma(r) =1,\nonumber\\
   &&V(r) =\Phi -\frac{Q}{r}\  _2F_1 \left[\frac{1}{4}, \frac{1}{2}, \frac{5}{4}, -\frac{b Q^2}{r^4}\right],
   \end{eqnarray}
   where the chemical potential $\Phi = \frac{Q}{r_h}\  _2F_1 \left[\frac{1}{4}, \frac{1}{2}, \frac{5}{4}, -\frac{b Q^2}{r_h^4}\right]$ so that the gauge condition $V(r_h)=0$ is satisfied. And $_2F_1$ is the hypergeometric function. In the limit $b \rightarrow 0$, it reduces to the well-known RN BH.
   
   Now the problem reduces to solving the scalar field equation (\ref{EOMs}) on the fixed Born-Infeld BH, which can be interpreted as that of a scalar field in an effective potential $U_{\textrm{eff}} = \mu^2_{\textrm{eff}} \psi^2 - \lambda \psi^4 + \nu \psi^6$ with $\mu^2_{\textrm{eff}}  \equiv \mu^2 - \frac{q^2 V^2}{N \sigma^2}$. As the Maxwell case with $b=0$ discussed in Refs. \cite{Brihaye:2020vce,Brihaye:2021phs,Garcia:2021pzd}, existence of scalar clouds requires the Coulomb potential energy $q \Phi$ to be sufficiently large while satisfying the bound state condition (\ref{BoundStateCondition}), i.e.,
   \begin{eqnarray}
   \sqrt{1 - \frac{\lambda^2}{4 \mu^2 \nu}} \leq q \Phi/\mu \leq 1,
   \end{eqnarray}
   where the lower bound is approximately derived by analyzing the asymptotic behavior of the effective potential $U_{\textrm{eff}}$ and confirmed by numerical calculations. For fixed chemical potential $\Phi$, the above condition implies that the gauge coupling constant $q$ should lie in a certain range, i.e. $q \in [q_{\textrm{min}} = \frac{\mu}{\Phi}\sqrt{1 - \frac{\lambda^2}{4 \mu^2 \nu}}, q_{\textrm{max}} = \frac{\mu}{\Phi}] $, for the possible existence of scalar clouds.
   
\begin{figure}[!htbp] 
     	\centering
     	\includegraphics[width=0.45\textwidth]{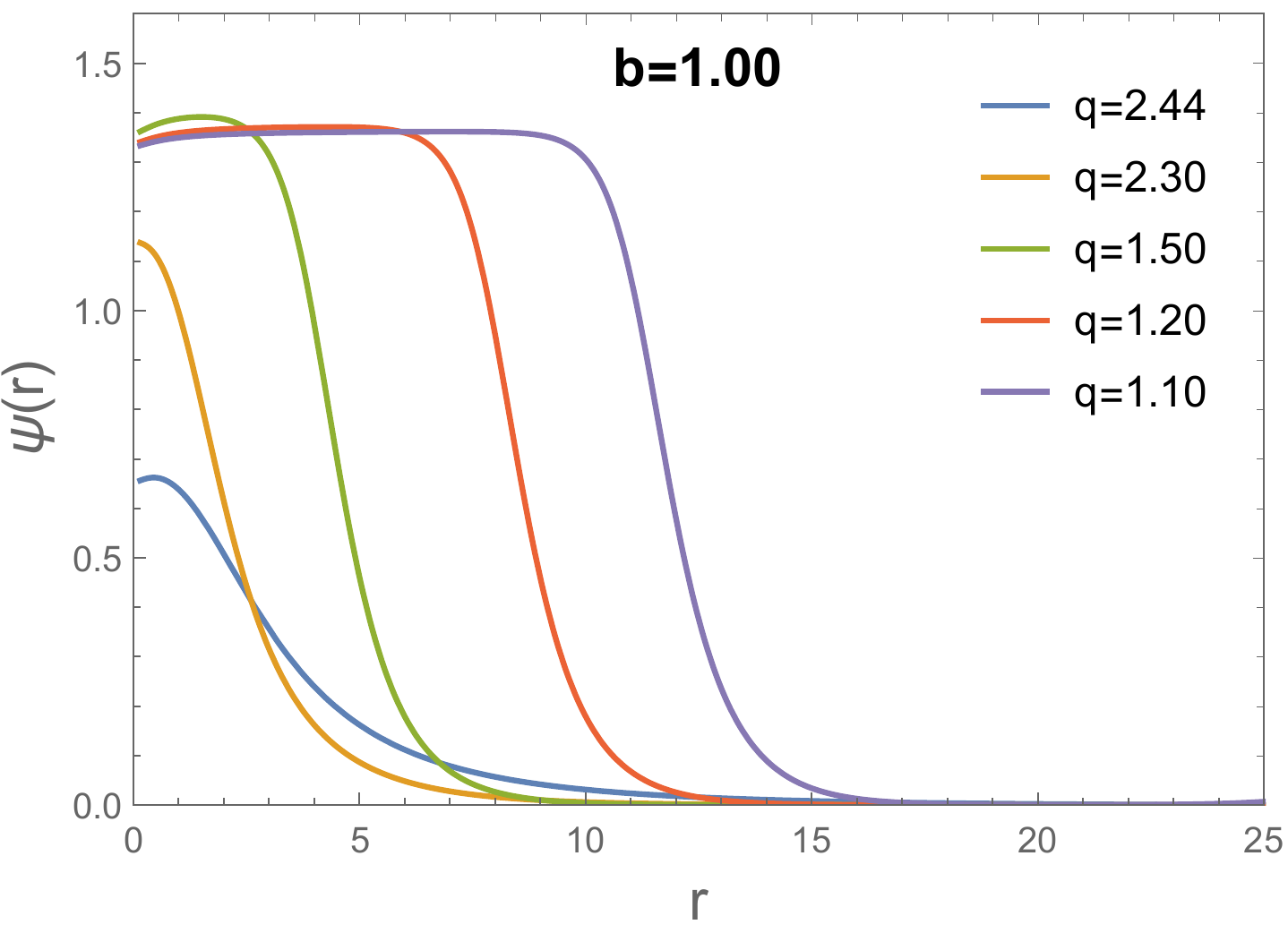}\quad
     	\includegraphics[width=0.45\textwidth]{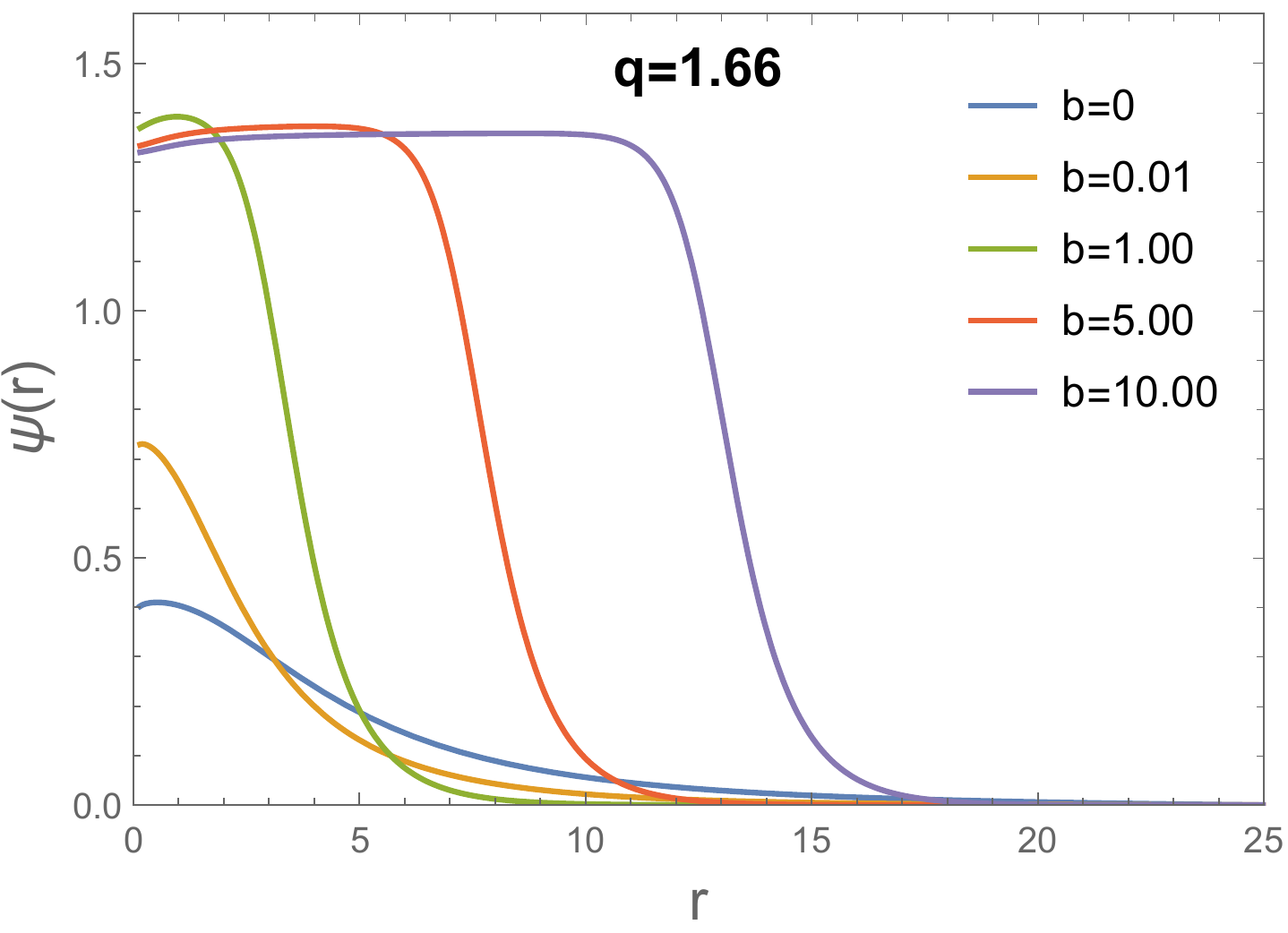}
     	\caption{Profiles of scalar clouds on Born-Infeld BHs for various values of $b$ and $q$. In the left panel, $b=1.00$ while $q$ varies. In the right panel $q=1.66$ while $b$ varies. Other parameters are set as $\nu = 9/32, r_h=0.15, Q=0.09$.}\label{ScalarBIBH}
     \end{figure}

   In Fig. \ref{ScalarBIBH}, profiles of scalar clouds as a function of the radial coordinate for various values of $b$ and $q$ are plotted. In the left panel, we fix $b=1.00$ and vary $q$. Notice that $q=2.44 \approx q_{\textrm{max}}$. From the panel, we can see that with the decreasing of $q$, the cloud becomes more extended outside the horizon and approaches a step-function-like shape. This phenomenon is similar to the $b=0$ case discussed in \cite{Brihaye:2020vce,Brihaye:2021phs,Garcia:2021pzd}. In the right panel, we fix $q=1.66$ which is approximately the value of $q_{\textrm{max}}$ for $b=0$ case. We vary the value of $b$ to see its effect on the scalar cloud. From the panel, we can see that the effect of $b$ is almost opposite to that of $q$: increasing $b$ will make the cloud more extended to take a step-function-like shape. This can be understood from the asymptotic behavior of the scalar field  Eq. (\ref{SpatialInfinity}). Increasing $q$ or decreasing $b$ will increase the asymptotic effective mass of the scalar field, and thus make the cloud more localized. Moreover, we should note that for fixed $q$, when $b$ is increased to large enough value, the cloud will cease to exist. 
    
   \subsection{Charged scalar clouds on Schwarzschild BHs}
   
   Now we consider the scalar field to be coupled with the gauge field, but they two are both decoupled from the metric. This can be achieved by taking the limit $\alpha \rightarrow 0$ ($G \rightarrow 0$ equivalently). The fixed background metric is the Schwarzchild BHs with
   \begin{eqnarray}
   N(r) = 1 - \frac{2 m}{r},\qquad \sigma(r) =1.
   \end{eqnarray}
   Then the problem reduces to solve the scalar field and gauge field equations (\ref{EOMs}) simultaneously on this fixed background.

   The mass of the matter fields outside the horizon can be evaluated by their Komar energy\cite{Herdeiro:2020xmb}
 \begin{eqnarray}
 M_Q = \frac{1}{4\pi} \int d^3 x \sqrt{-g} (T_\alpha^{~\alpha} - 2 T_t^{~t}) = \Phi Q_e + M_\psi, 
 \end{eqnarray}
 where $M_\psi = 2 \int_{r_h}^\infty r^2 \sigma \left[\frac{q^2 V^2 \psi^2}{N \sigma^2} - U(\psi)\right]$ is the mass of the scalar field.
 
  \begin{figure}[!htbp] 
  	\centering
  	\includegraphics[width=0.45\textwidth]{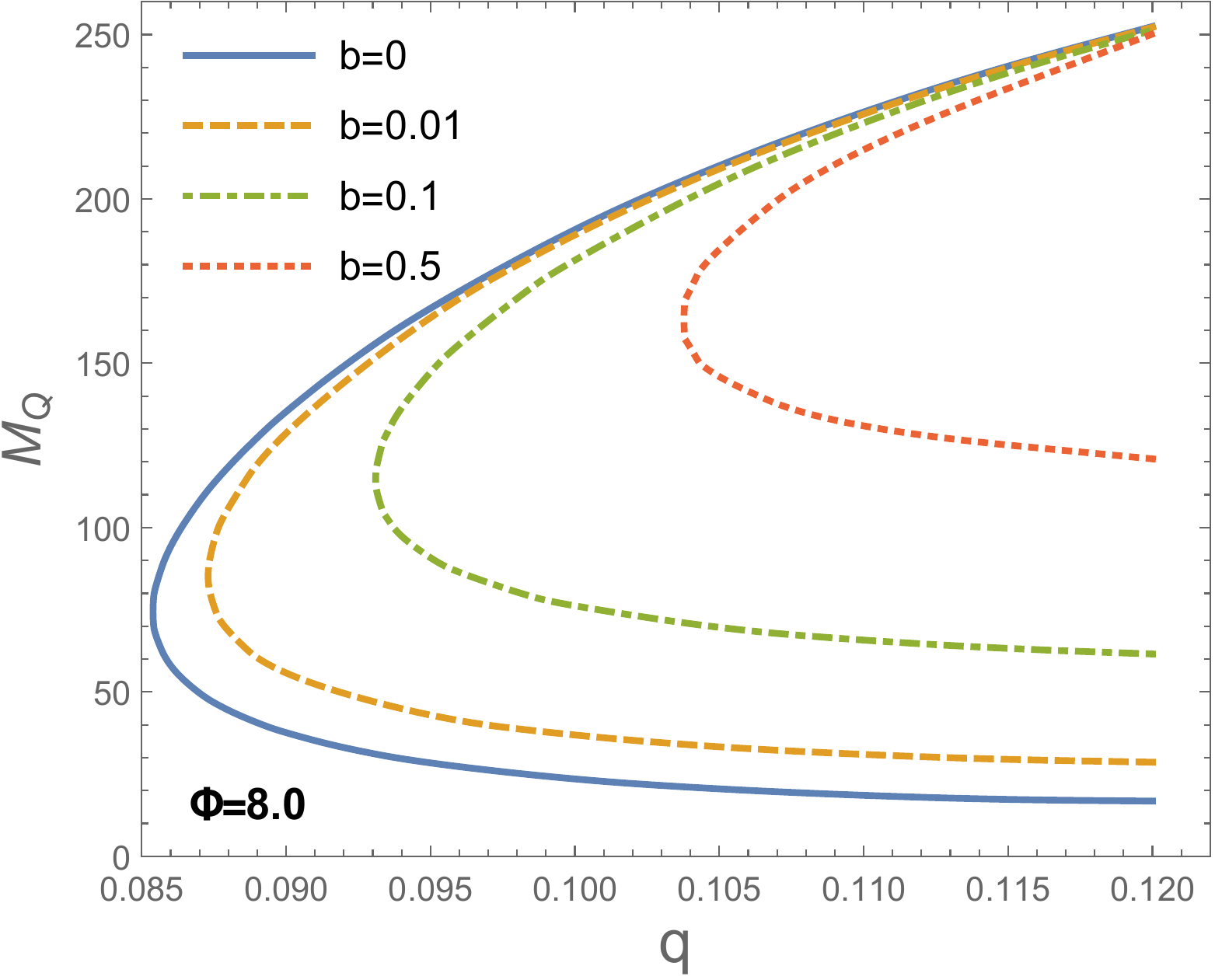}\quad
  	\includegraphics[width=0.45\textwidth]{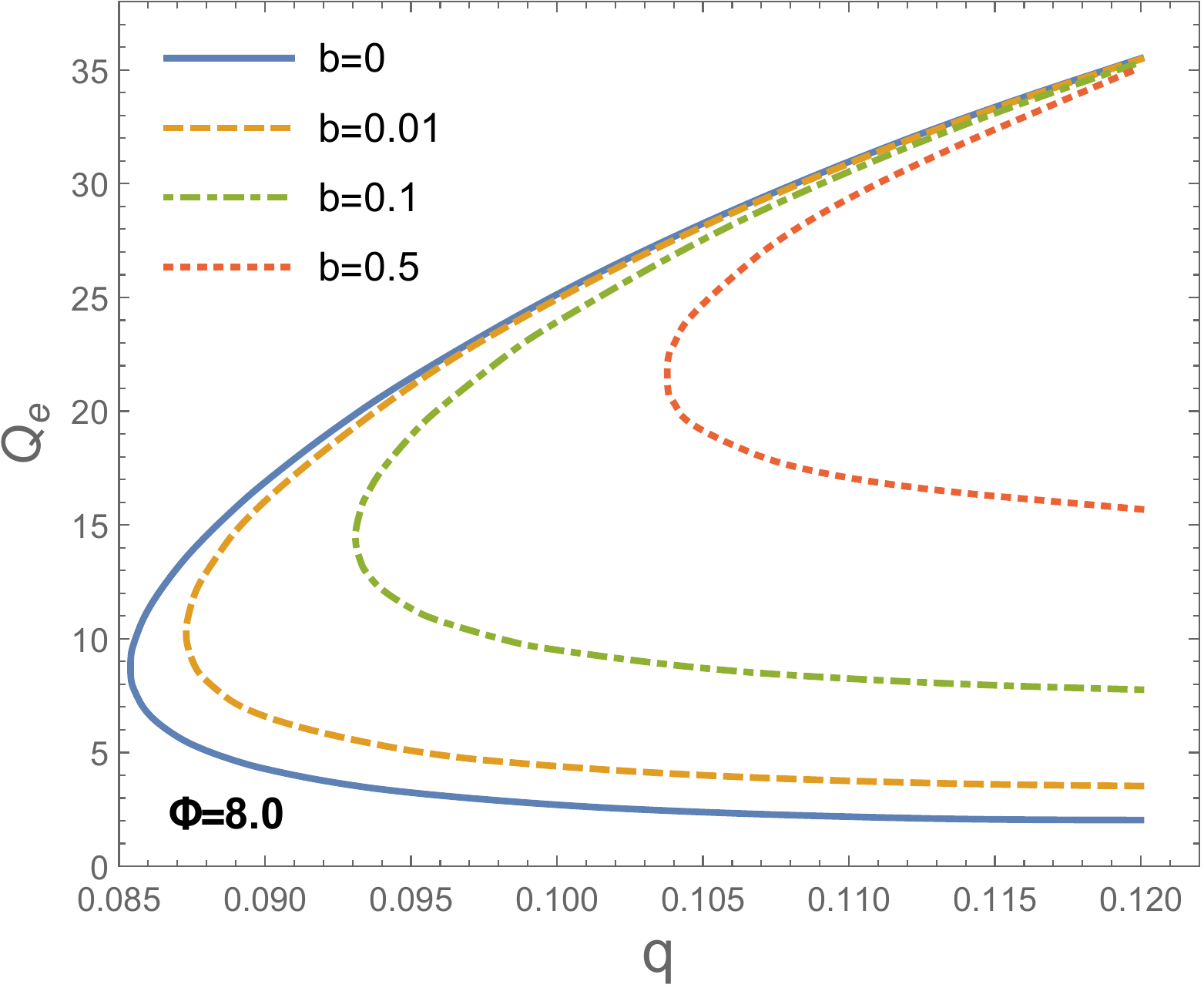}
  	\caption{Scalar clouds on Schwarzschild BHs. Dependence of the mass $M_Q$ (left-panel) and the electric charge $Q_e$ (right panel) on values of $q$ and $b$ for fixed chemical potential $\Phi=8.0$. Other parameters are set as $\nu = 9/32, r_h=0.15$. }\label{QCloudsSchw1}
  \end{figure}
 
Series of solutions are constructed numerically for various values of $q$ and $b$, and their mass $M_Q$ and electric charge $Q_e$ are computed, as shown in Fig. \ref{QCloudsSchw1}. From the figure, several silent properties can be observed:
 \begin{itemize}
 	\item For fixed $b$, as in the case of scalar clouds on the Born-Infeld BHs discussed above, the gauge coupling constant $q$ needs to lie in a certain range, $q \in [q_{\textrm{min}}, q_{\textrm{max}}]$, for the existence of the clouds on Schwarzschild BH. The upper bound $q_{\textrm{max} }= \frac{\mu}{\Phi}$ is once again from the bound state condition (\ref{BoundStateCondition}) while the lower bound $q_{\textrm{min}}$ can only determined numerically. Moreover, increasing $b$ will make the allowed range $q_{\textrm{max} } - q_{\textrm{min}}$ for the existence of the clouds shrink. And when $b$ exceeds some threshold value, the range $q_{\textrm{max} } - q_{\textrm{min}}$ disappears and the cloud will no longer exist.
 	
  	\item Similar to the Maxwell case \cite{Hong:2019mcj,Herdeiro:2020xmb,Hong:2020miv,Brihaye:2020vce,Garcia:2021pzd,Brihaye:2021phs}, there is a mass gap of the cloud: the mass/electric charge of the cloud can not take arbitrary small value, and thus the cloud should not emerge as zero mode. 
 	
 	\item For fixed $b$, there exist two branches of solutions, which can be named as branch A with smaller mass and branch B with larger mass respectively. At $q = q_{\textrm{min}}$, the two branches of solutions join. This phenomenon has been observed in the Maxwell case \cite{Hong:2019mcj,Herdeiro:2020xmb,Hong:2020miv,Brihaye:2020vce,Garcia:2021pzd,Brihaye:2021phs}. The Born-Infeld coupling constant $b$ has rather different influences on the two branches of solutions. For branch A solutions, increasing $b$ will increase $M_Q$ and $Q_e$; While for branch B solutions, we have the opposite behaviors.

 \end{itemize} 
 
    \begin{figure}[!htbp] 
    	\centering
    	\includegraphics[width=0.45\textwidth]{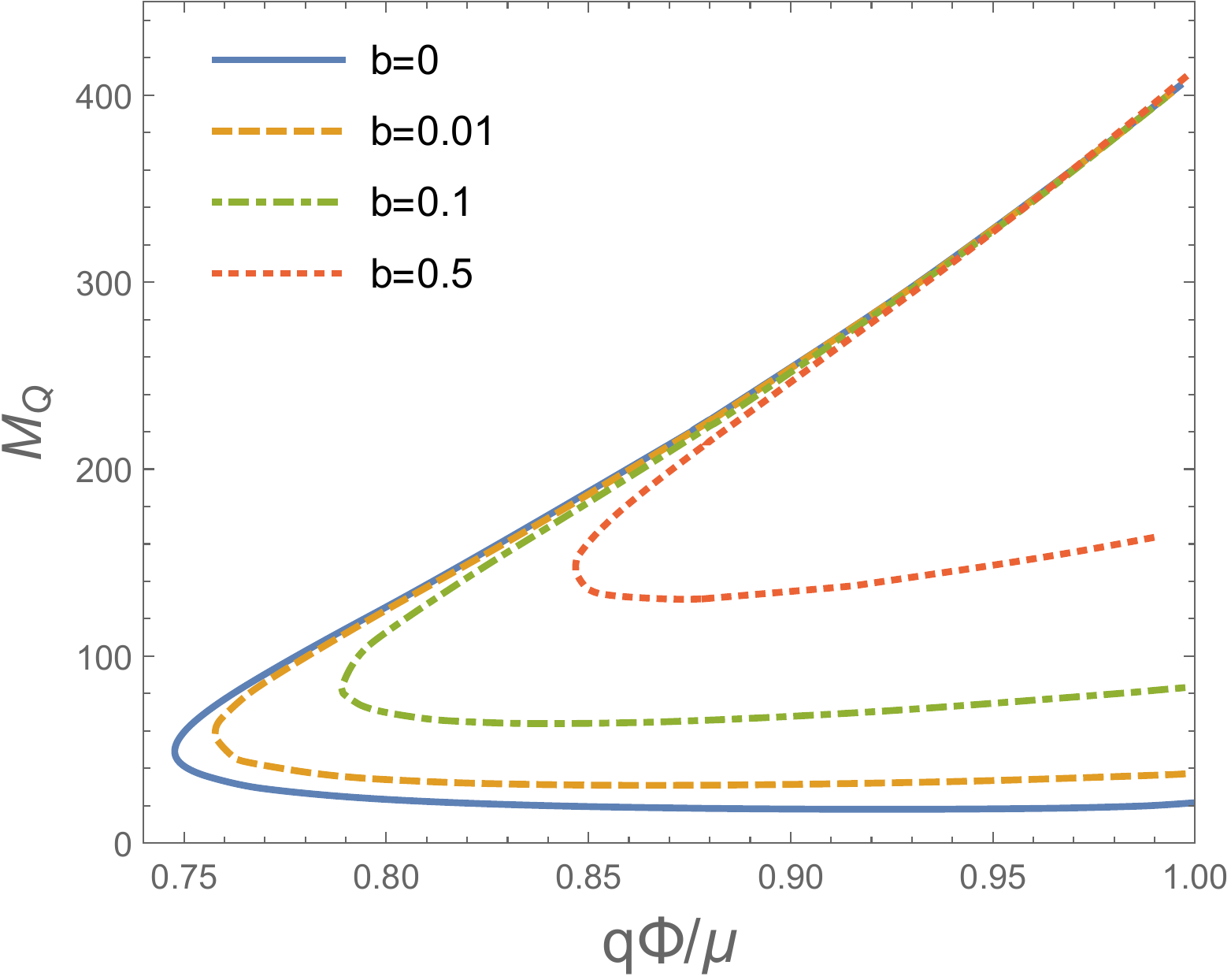}\quad
    	\includegraphics[width=0.45\textwidth]{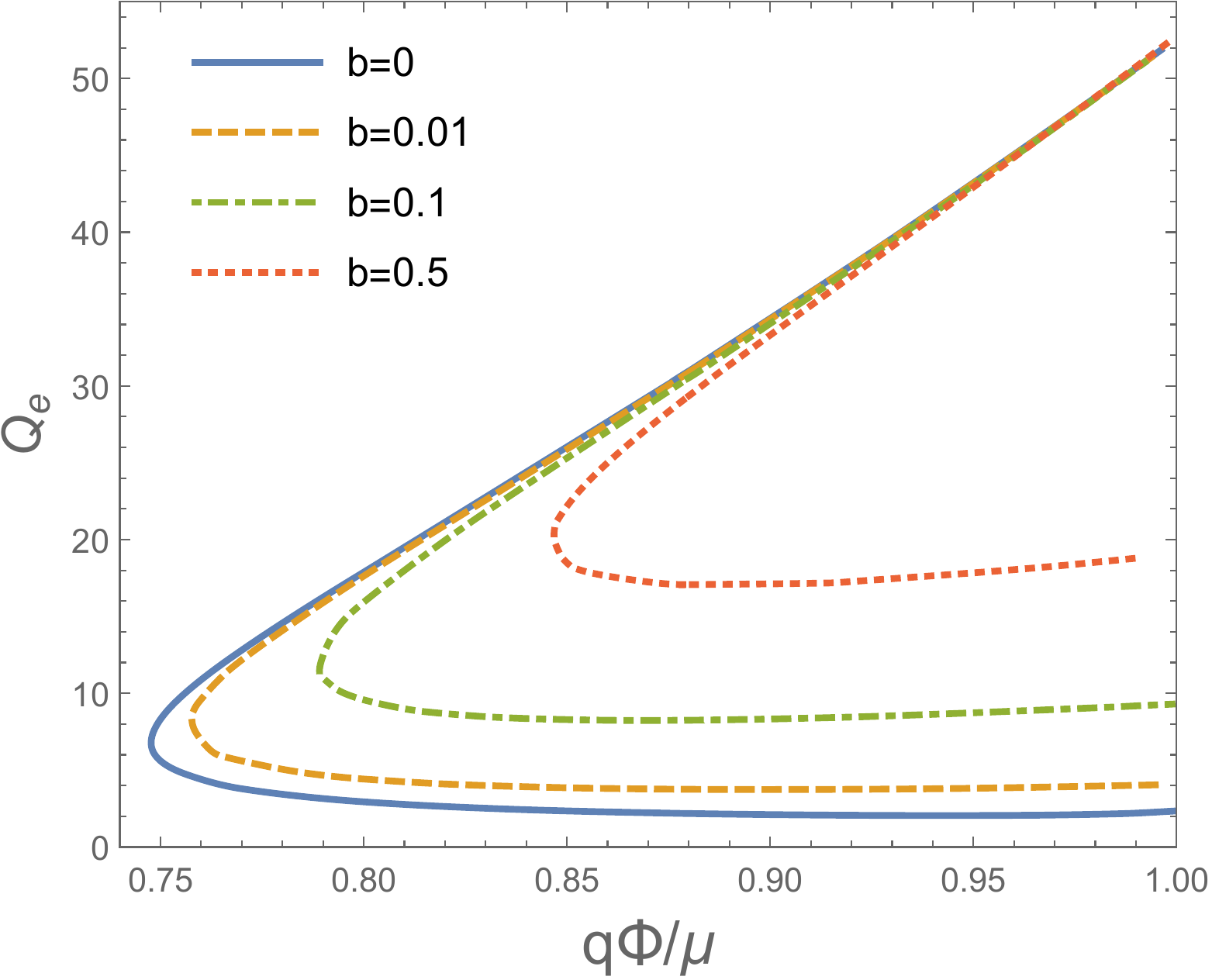}
    	\caption{Scalar clouds on Schwarzschild BHs. Dependence of energy $M_Q$ (left-panel) and the electric charge $Q_e$ (right panel) on values of $\Phi$ and $b$ for fixed gauge coupling constant $q=0.11$. Other parameters are set as $\nu = 9/32, r_h=0.15$.}\label{QCloudsSchw2}
    \end{figure}

 The above effects of $b$ on the clouds can also be observed from Fig. \ref{QCloudsSchw2} where series of solutions are constructed with fixed $q$ but varied $\Phi$. In this time, scalar cloud can only exist for $\Phi \in [\Phi_{\textrm{min}}, \Phi_{\textrm{max}}]$, where the upper bound $\Phi_{\textrm{max}} = \frac{\mu}{q}$ is once again imposed by the bound state condition Eq. (\ref{BoundStateCondition}) while the lower bound $\Phi_{\textrm{min}}$ can only determined numerically. Increasing $b$ will increase $\Phi_{\textrm{min}}$ and thus the existence domain of the cloud shrinks. And when $b$ is large enough to exceed some threshold value, the existence domain disappears and the cloud will no longer exist. Also, from the physical quantities of the cloud $M_Q$ and $Q_e$, one can see that the influences of $b$ on the two branches of solutions are significantly different.
 
  \begin{figure}[!htbp] 
  	\centering
  	\includegraphics[width=0.45\textwidth]{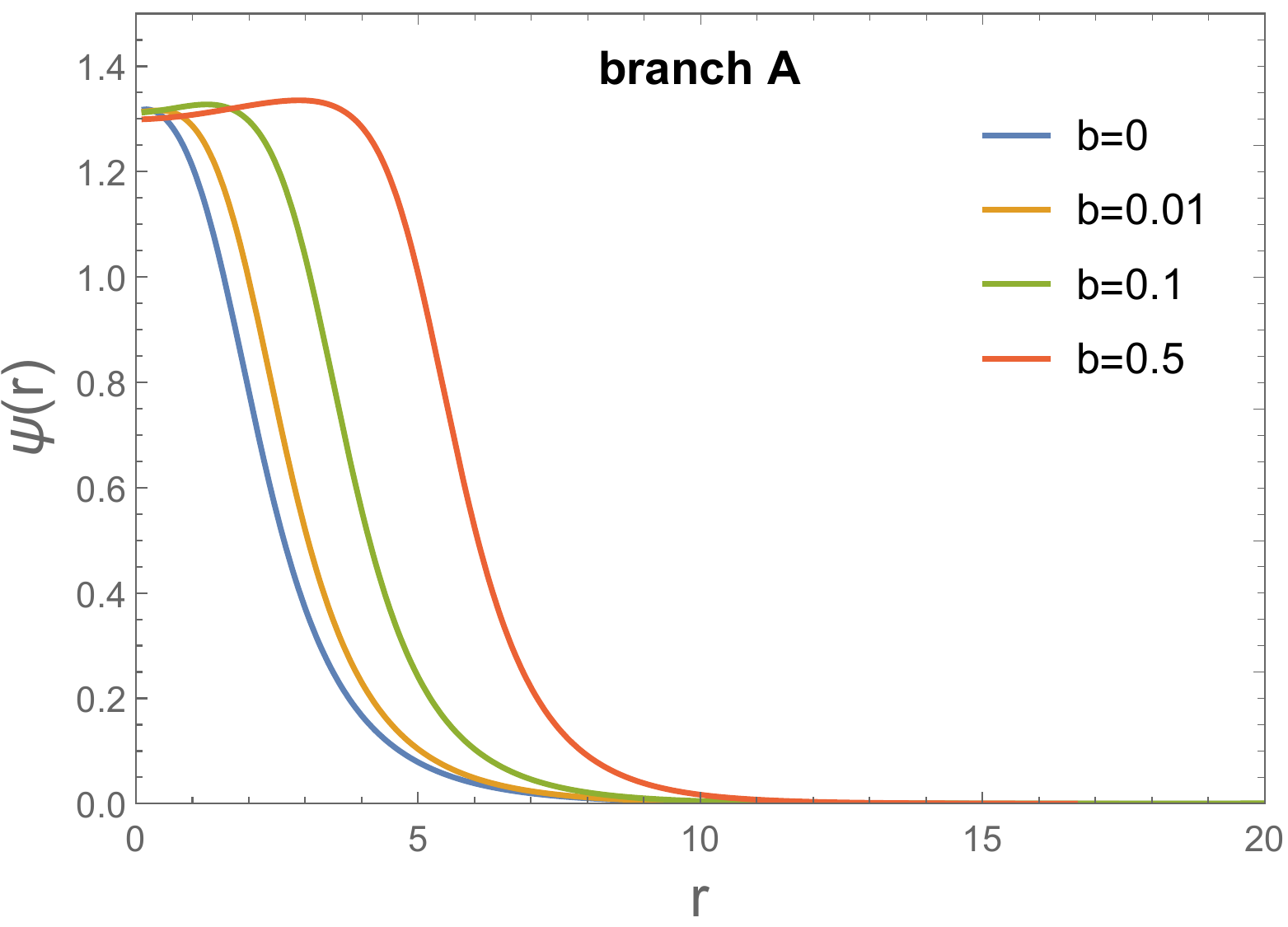}\quad
  	\includegraphics[width=0.45\textwidth]{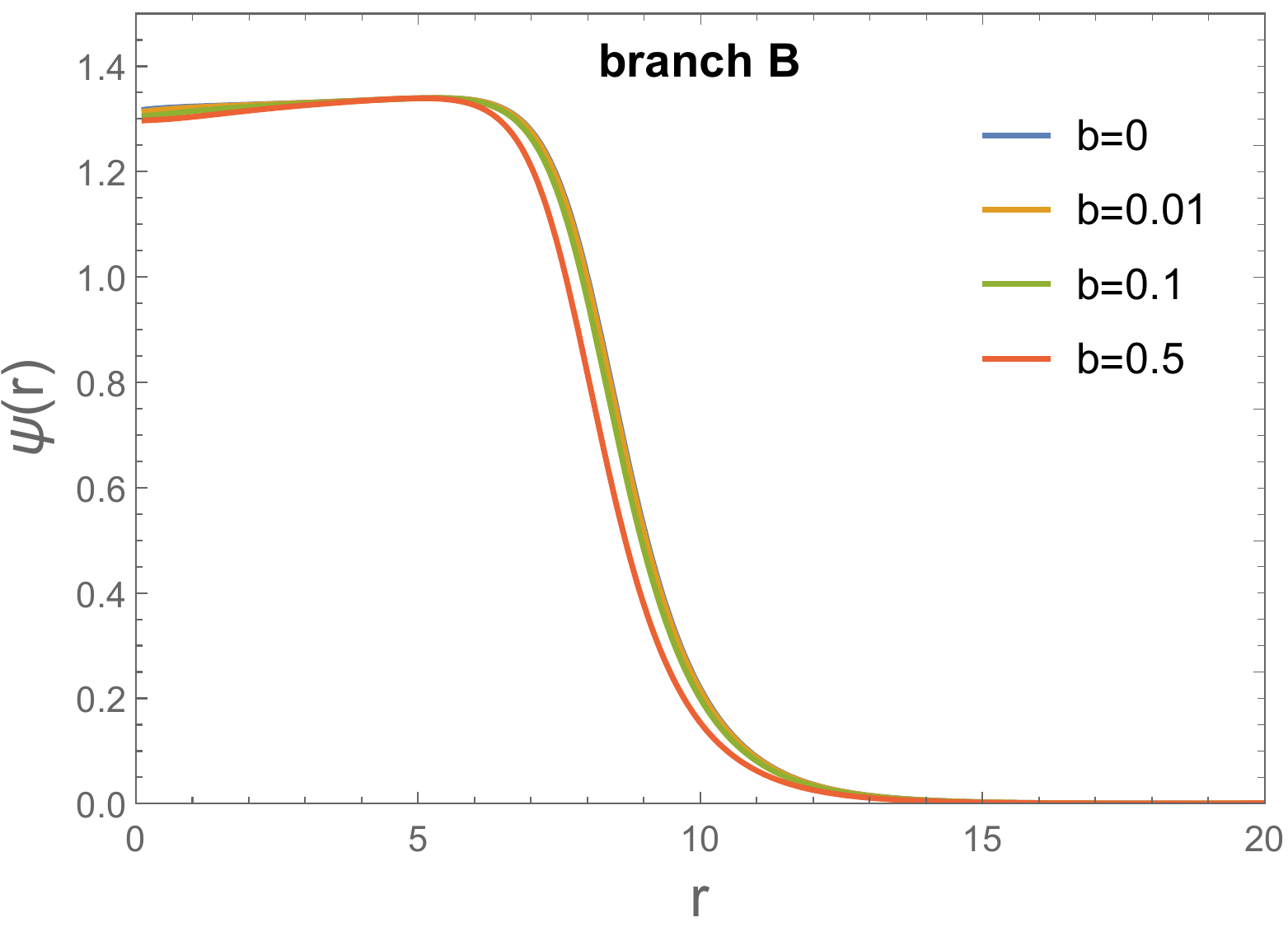}
  	\caption{Typical profiles of the two branches of scalar clouds on Schwarzschild BHs for fixed chemical potential $\Phi=8.0$ and gauge coupling constant $q=0.11$. Other parameters are set as $\nu = 9/32, r_h=0.15$.}\label{QCloudsSchw3}
  \end{figure}

 In Fig. \ref{QCloudsSchw3}, typical profiles of the scalar field are plotted for the two branches of solutions. From the figure, one can see the two branches of clouds take different configurations with the branch B clouds being more likely to take a step-function-like shape. For branch A clouds, increasing $b$ will make the scalar field configuration more extended to take a step-function-like shape; While for the branch B clouds, $b$ has little effect on their shape.

\subsection{Charged spherical BHs with charged scalar hair}

\begin{figure}[!htbp] 
	\centering
	\includegraphics[width=0.31\textwidth]{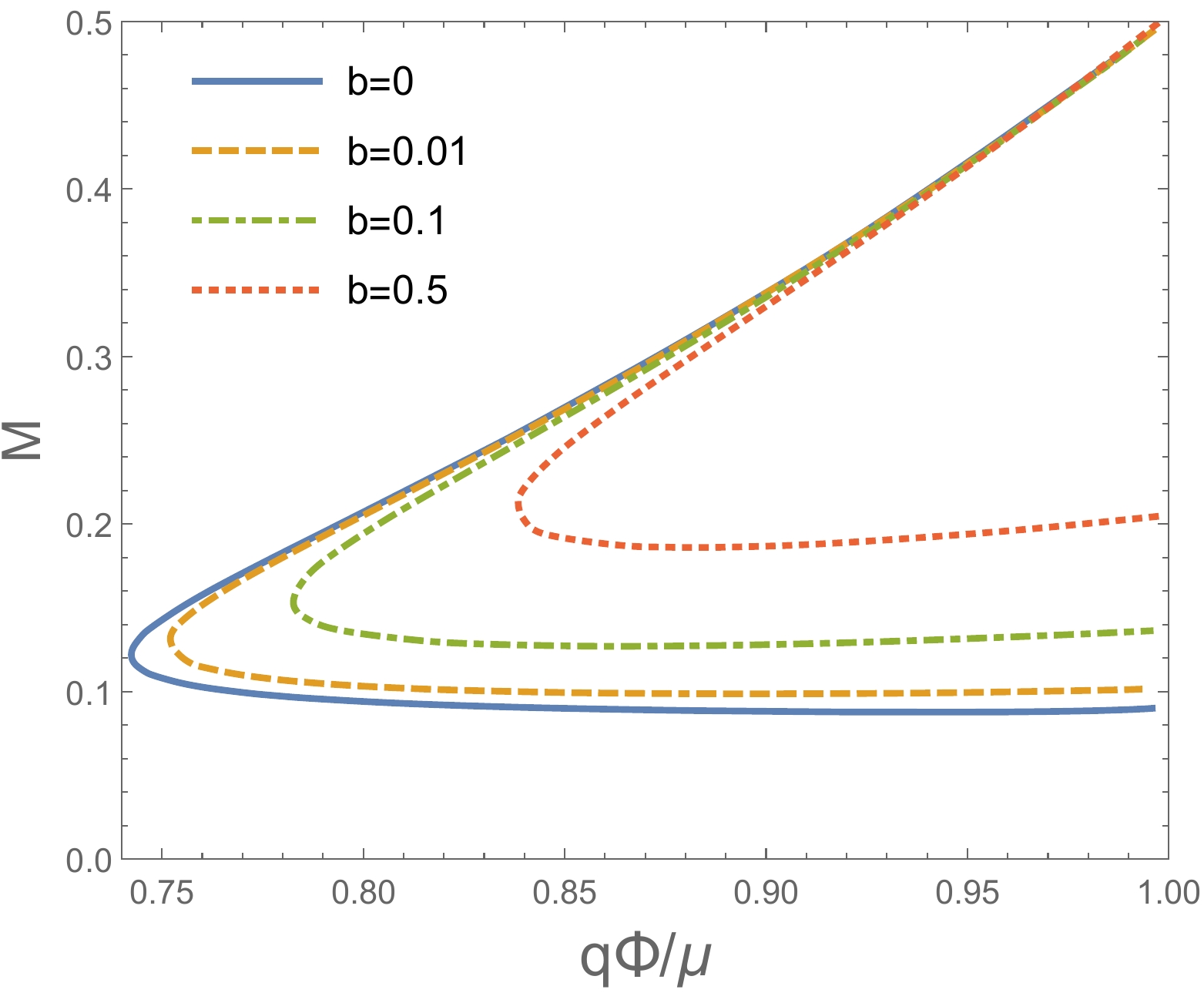}\quad
	\includegraphics[width=0.31\textwidth]{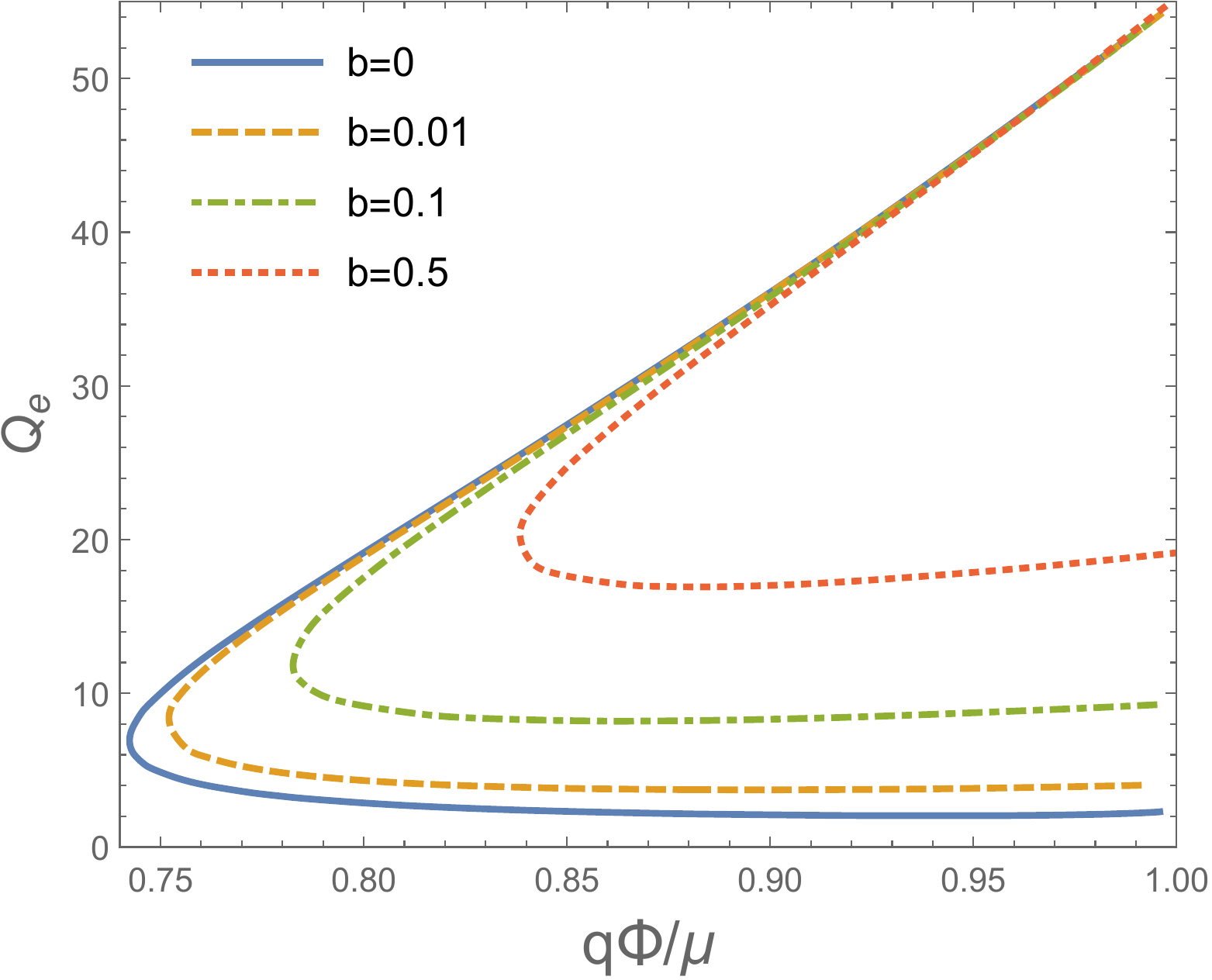}\quad
	\includegraphics[width=0.31\textwidth]{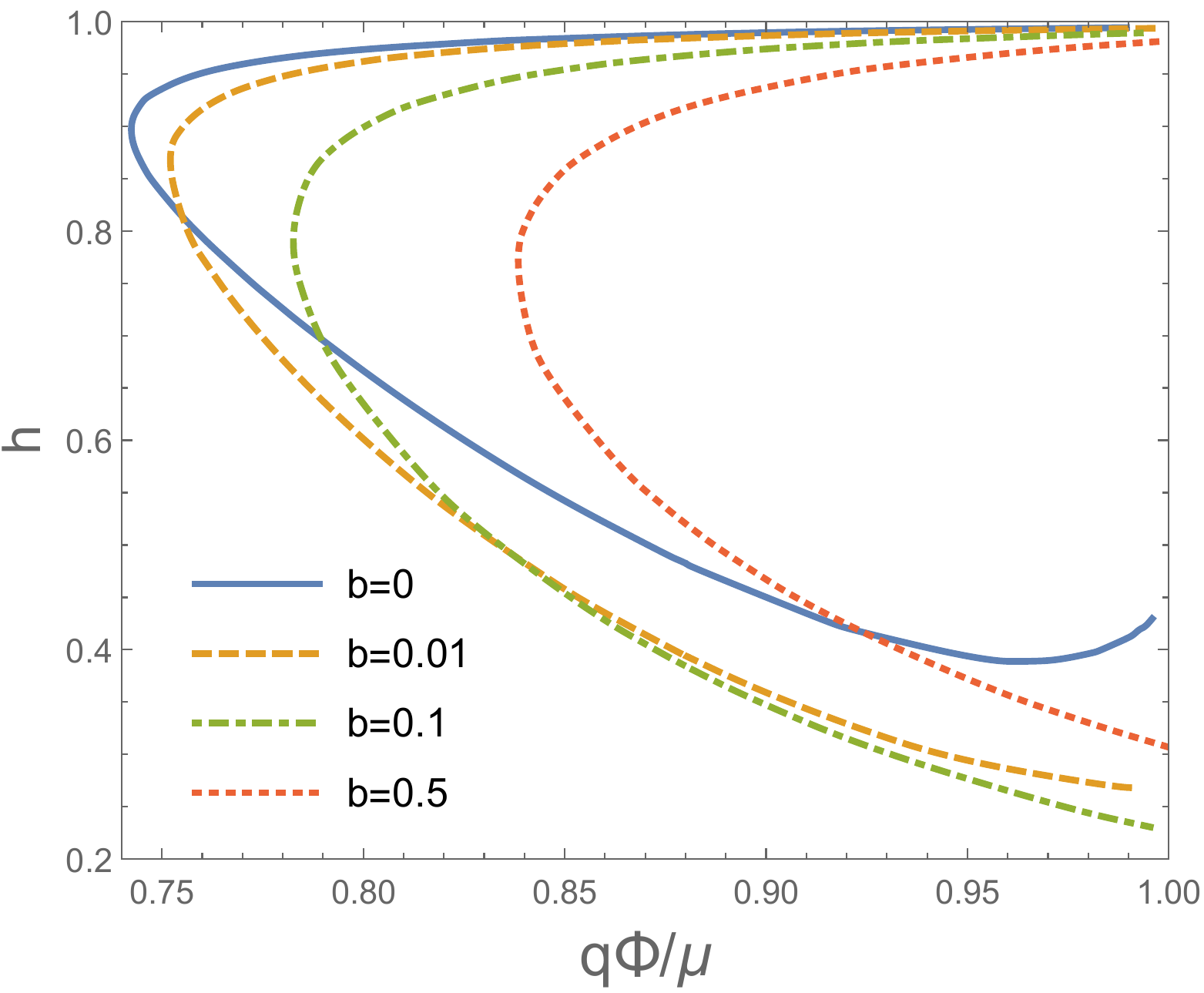}
	\caption{Charged spherical BHs with charged scalar hair. Dependence of the ADM mass $M$, electric charge $Q_e$ and hairiness $h$ on values of $b$ and the chemical potential $\Phi$ for fixed charge $q=0.11$ and gravitational coupling constant $\alpha=0.001$. Other parameters are set as $\nu = 9/32, r_h=0.15$.}\label{QHair1}
\end{figure}

Now we turn on the gravitational coupling constant $\alpha$, then the clouds will have back-reaction on the geometry to form hairy BHs. Series of hairy BH solutions for various values of $\alpha, b$ and $q$ are constructed numerically. The ADM mass $M$, the electric charge $Q_e$ and the hairiness $h$ of these solutions are thus computed. As an example, we show a sequence of solutions in Fig. \ref{QHair1} where $\alpha=0.001$ and $q=0.11$, from which one can see that the effects of $b$ is similar to that in the case $\alpha \rightarrow 0$ discussed above. Once again, increasing $b$ will make the existence domain of hairy BH solutions shrink. And when $b$ is large enough to exceed some threshold value, the existence domain disappears and hairy solutions will no longer exist. Moreover, from the figure, one can see that $Q_e$ and $h$ can not take arbitrary small values which means that the hairy solutions can not degenerate to the bald ones (Schwarzschild or Born-Infeld BHs) by reducing the amount of hair. Moreover, for branch B solutions, increasing $b$ will decrease the hairiness $h$; While for branch A solutions, influence of $b$ on the hairiness becomes complicated.

\begin{figure}[!htbp] 
	\centering
	\includegraphics[width=0.45\textwidth]{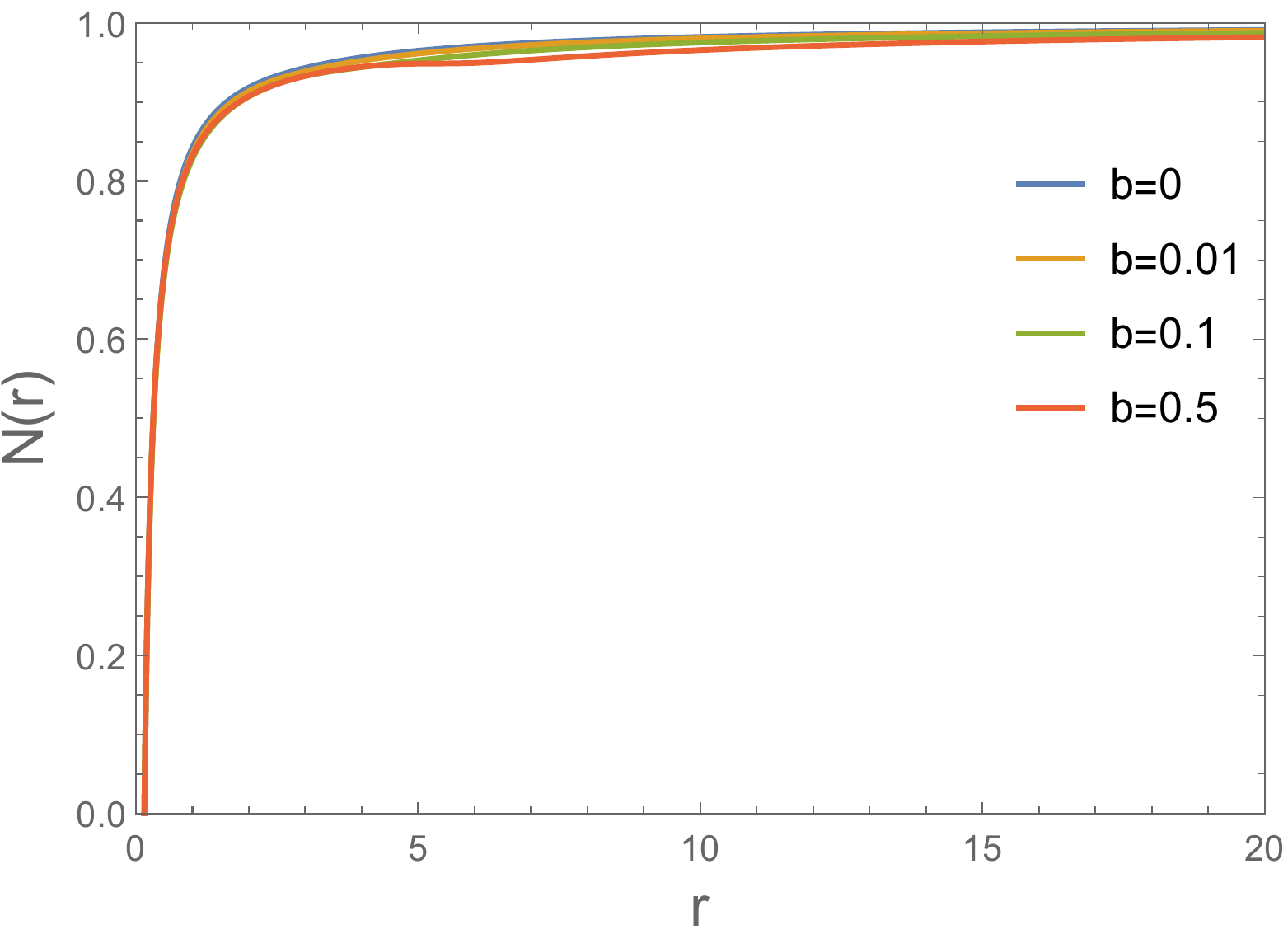}\quad
	\includegraphics[width=0.45\textwidth]{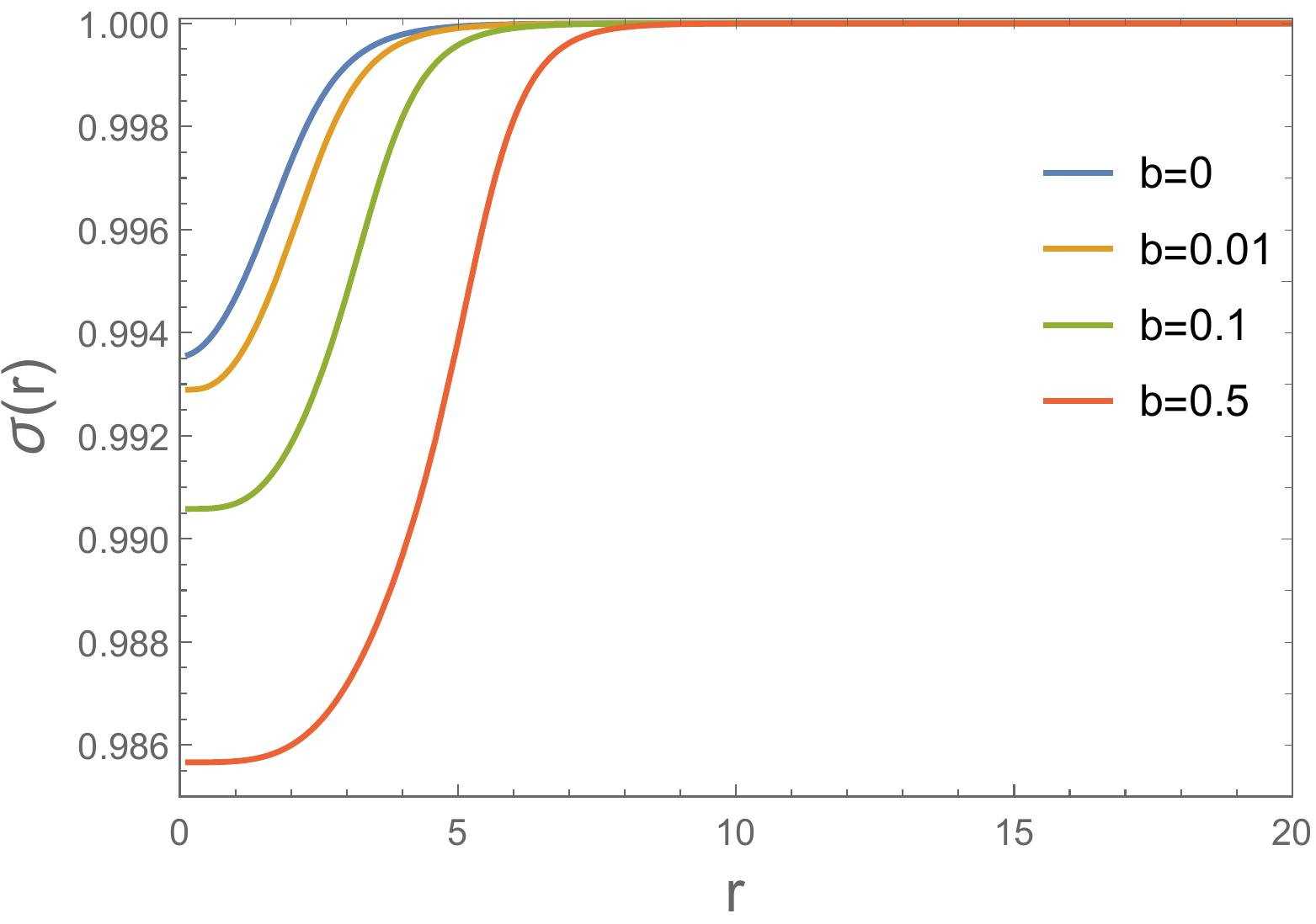}
	\includegraphics[width=0.45\textwidth]{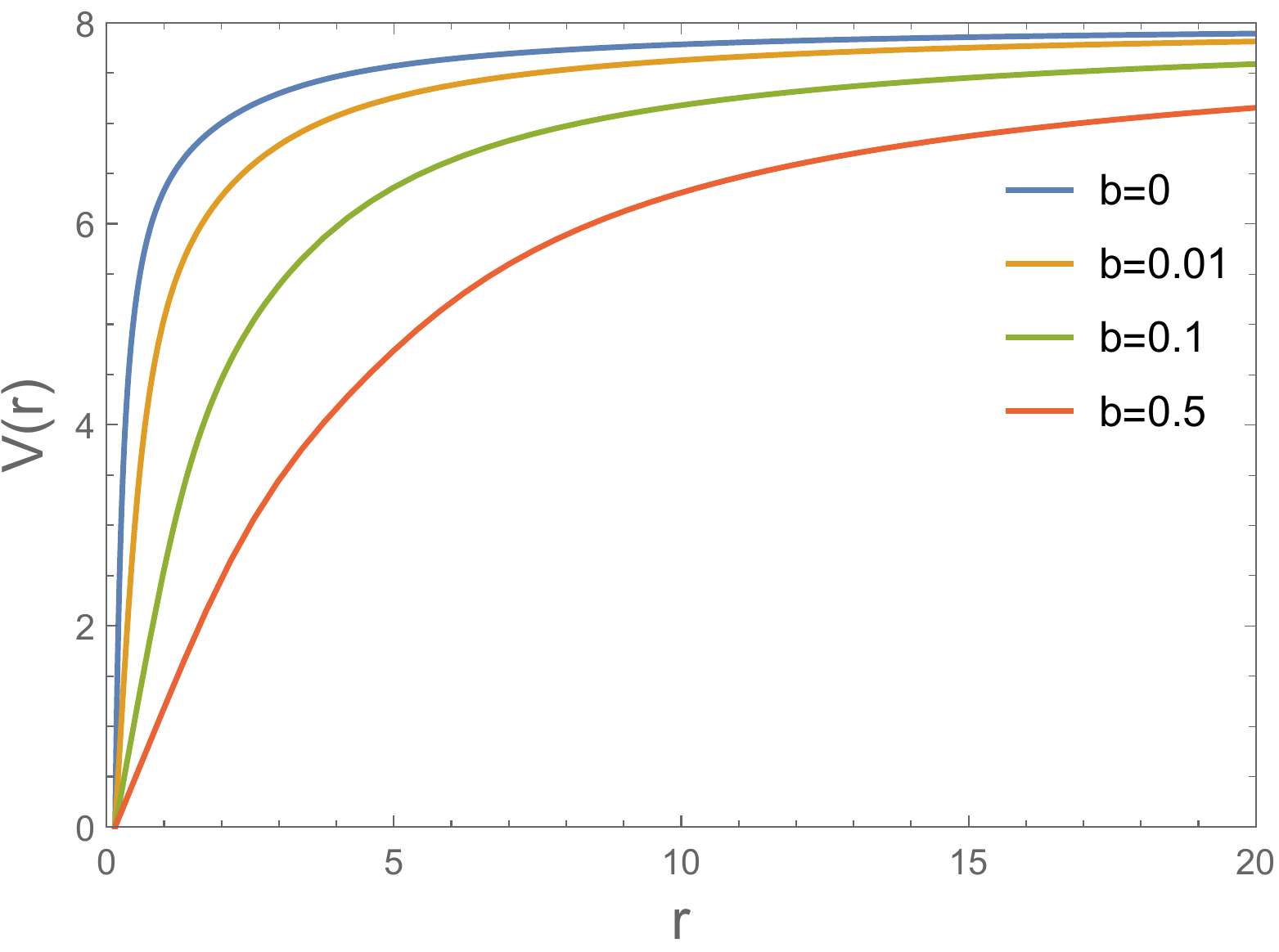}\quad
	\includegraphics[width=0.45\textwidth]{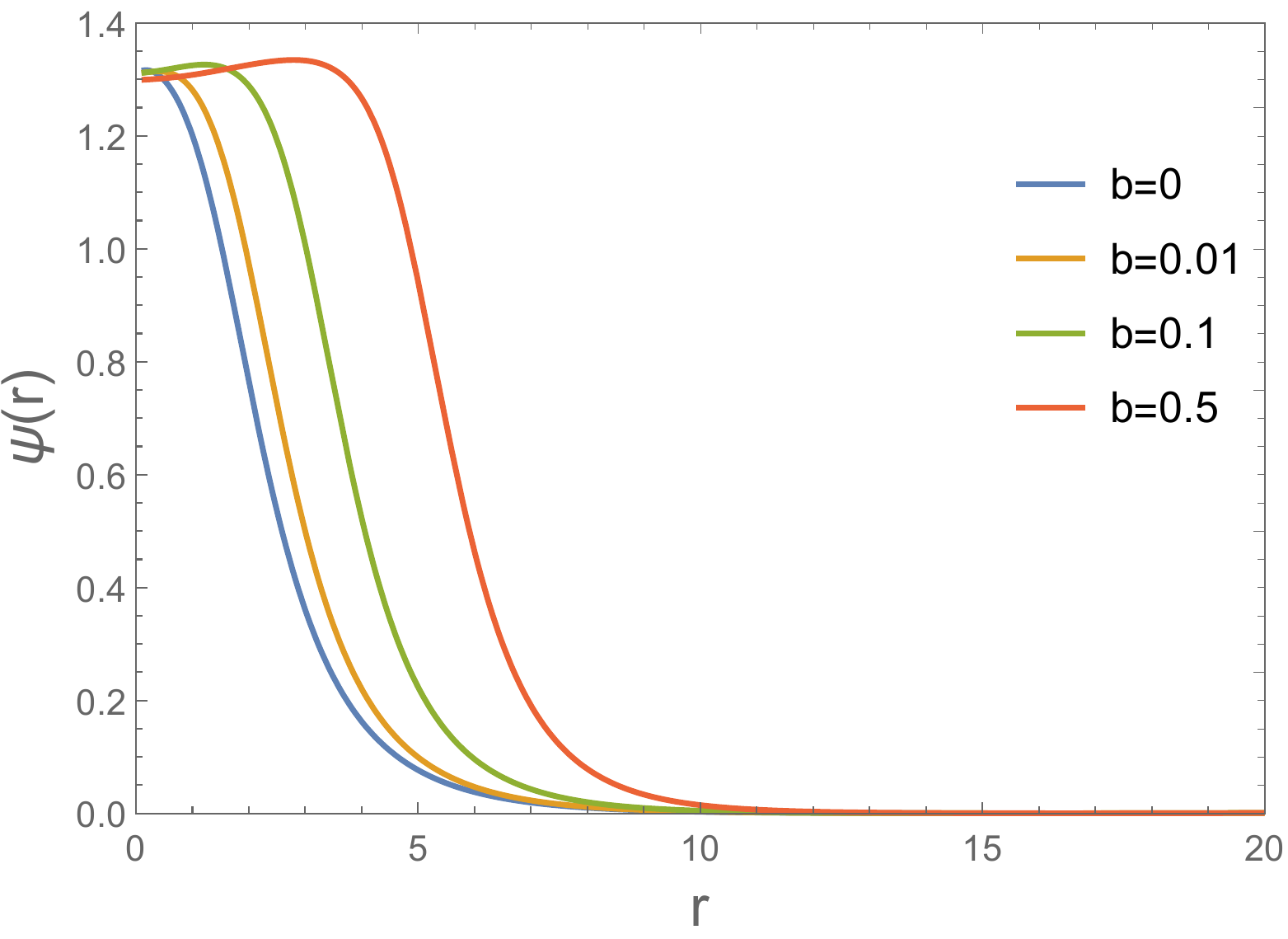}
	\caption{Profile functions of BHs with scalar hair for various $b$ and fixed chemical potential $\Phi=8.0$. Other parameters are set as $\alpha=0.001, q=0.11, \nu = 9/32, r_h=0.15$.}\label{QHair2}
\end{figure}

In Fig. \ref{QHair2}, as an illustrative example, we choose $\Phi=8.0$ and plot the profile functions of branch A solutions. One can see that with the increase of $b$, the Coulomb potential $V(r)$ becomes less steep near the horizon and the scalar hair $\psi(r)$ becomes more extended outside the horizon.

\begin{figure}[!htbp] 
	\centering
	\includegraphics[width=0.31\textwidth]{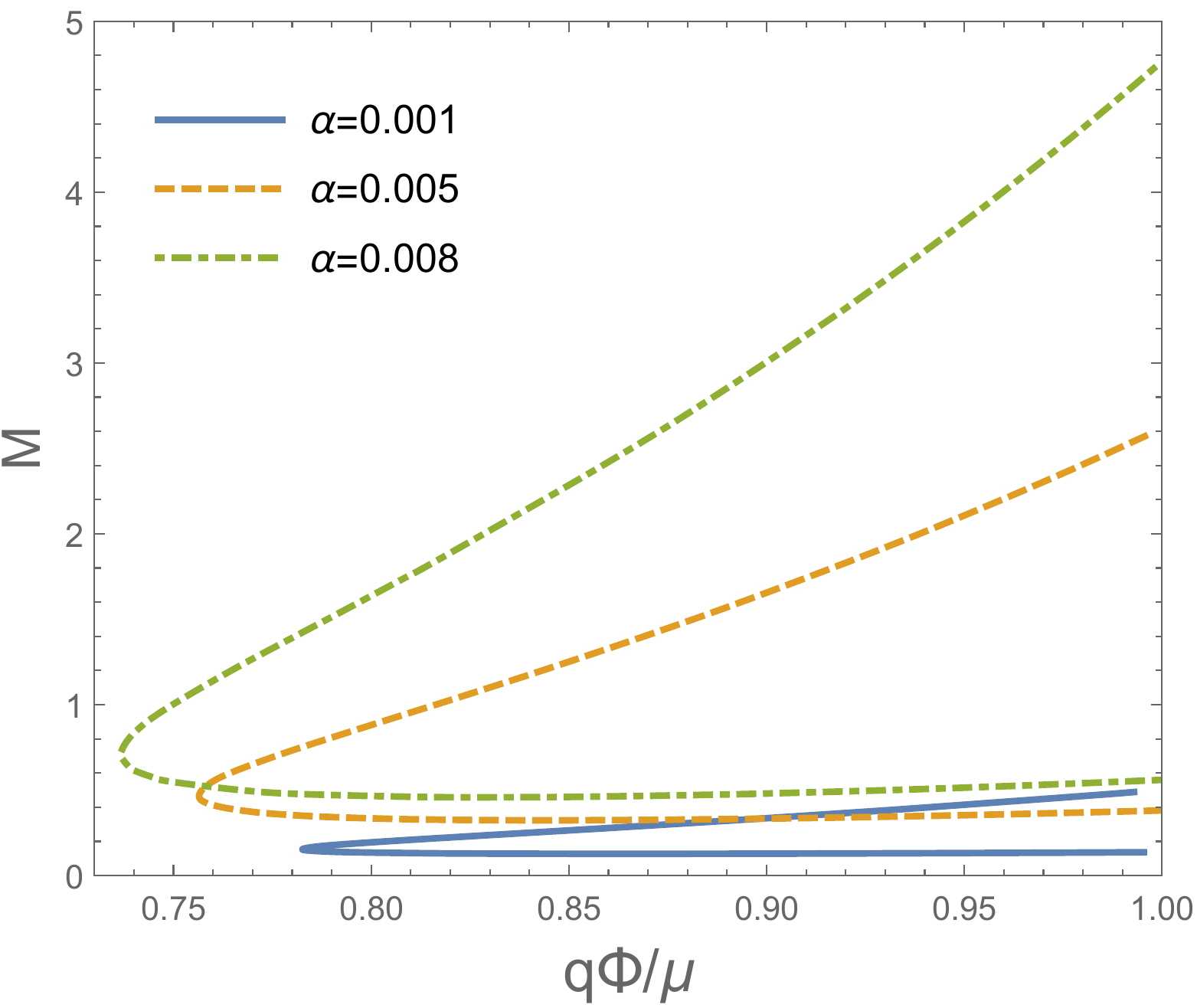}\quad
	\includegraphics[width=0.31\textwidth]{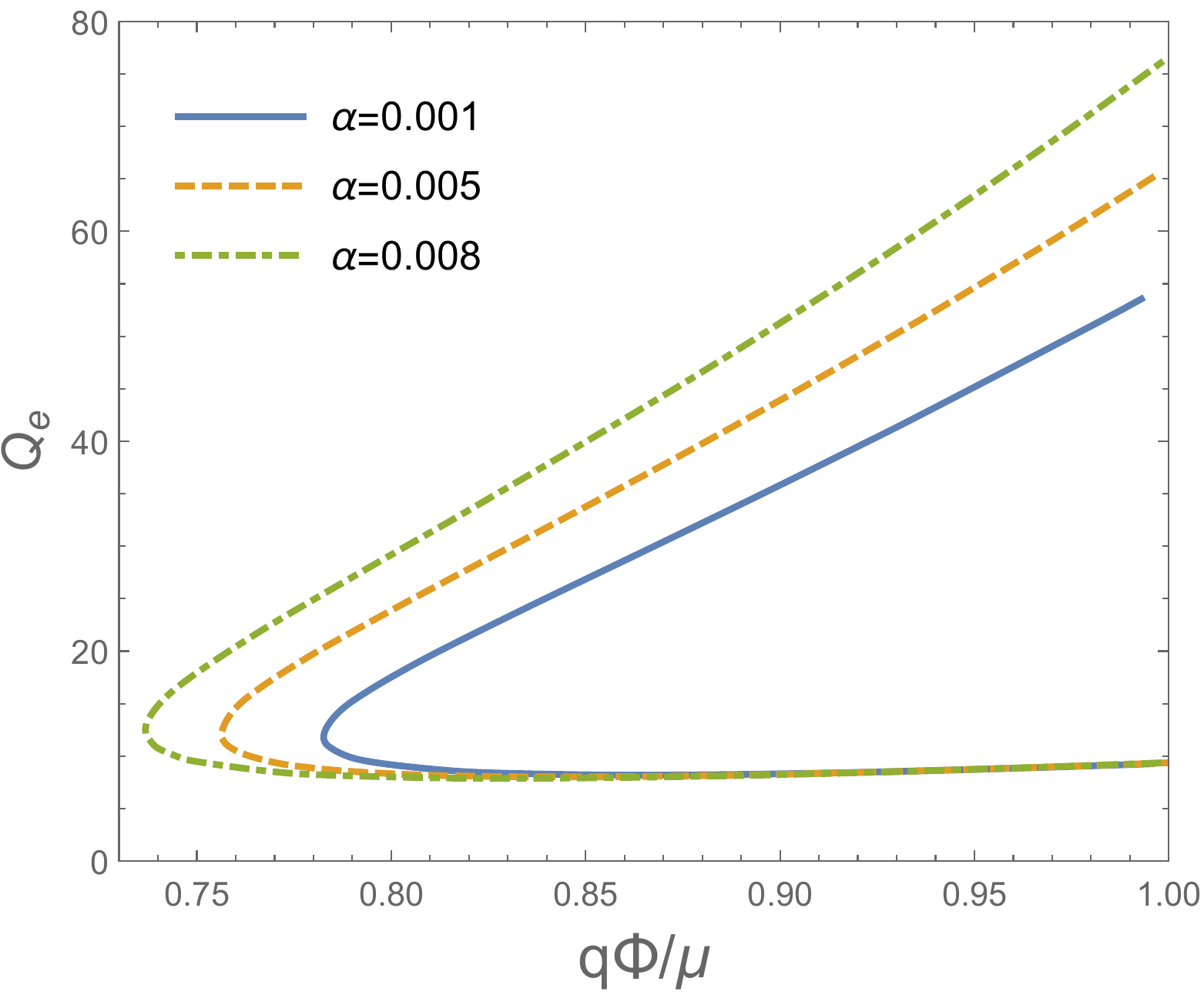}\quad
	\includegraphics[width=0.31\textwidth]{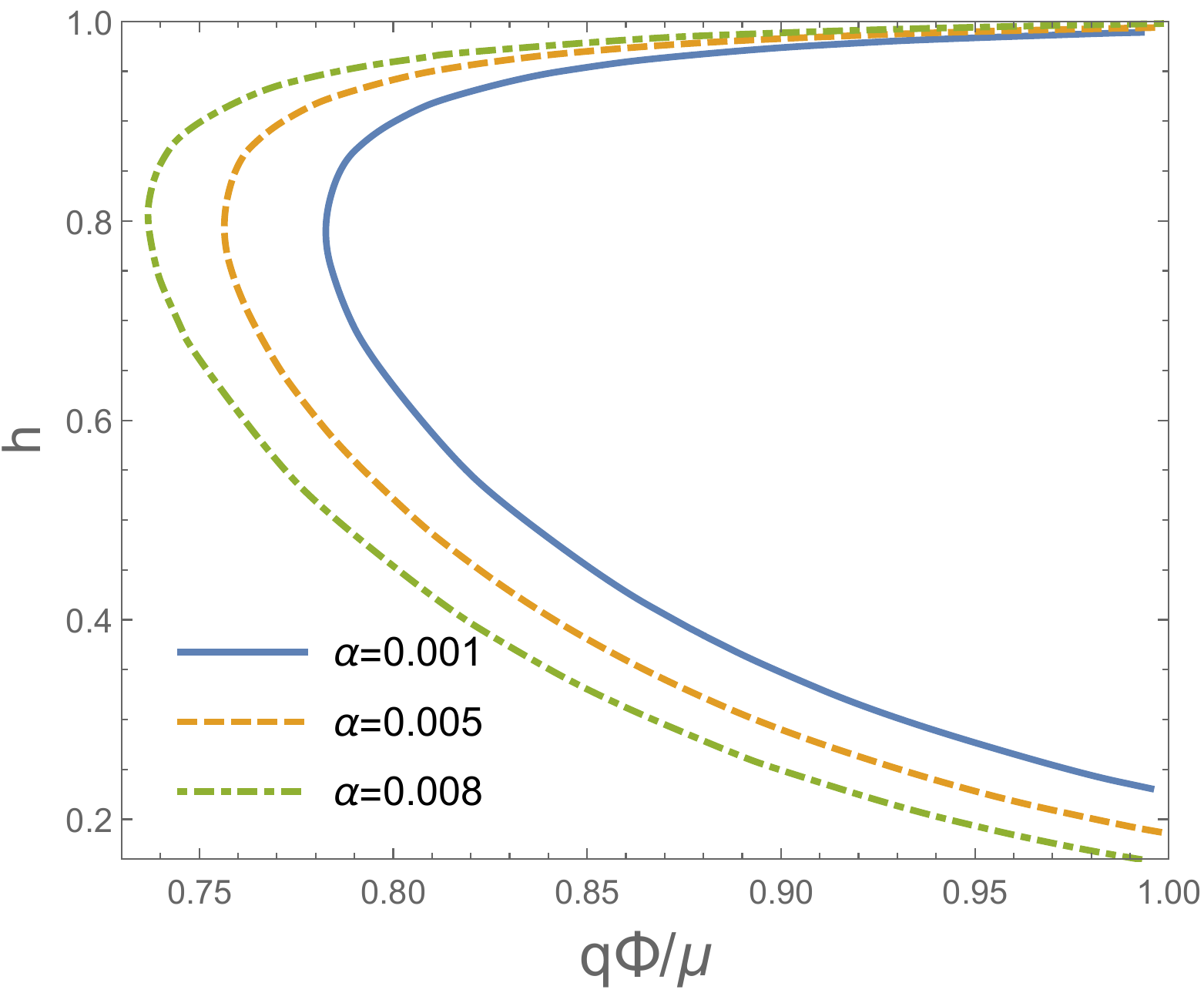}
	\caption{Charged BHs with charged scalar hair. Dependence of the ADM mass $M$, electric charge $Q_e$ and hairiness $h$ on values of the chemical potential $\Phi$ and $\alpha$ for fixed charge $q=0.11$ and Born-Infeld coupling constant $b=0.1$. Other parameters are set as $\nu = 9/32, r_h=0.15$. }\label{QHair3}
\end{figure}

To see the effects of turning on the gravitational coupling constant $\alpha$, sequences of solutions are constructed for fixed $b$ but varied $\alpha$, as shown in Fig. \ref{QHair3}. From the figure, one can see that, opposite to the effect of $b$,  increasing the gravitational coupling constant $\alpha$ will make the existence domain of hairy solutions enlarged. This implies that there is competition between the non-linearity of electromagnetic field and the gravitation on the hair formation. Moreover, $\alpha$ has significant influences on the physical quantities of the solutions.
 
\section{Summary and Discussions}

In this work, we construct asymptotically flat and spherical BHs with minimally coupled scalar cloud/hair in Einstein-Born-Infeld gravity. The scalar field takes a self-interaction potential which is essential for the existence of solutions. The cloud/hair endowed is charged and regular on the horizon. The metric remains static and spherical symmetric while the scalar field takes a harmonic time-dependence form $\Psi \sim \exp(i \omega t)$. Synchronization condition $\omega =\omega_c$ is required for the existence of scalar cloud/hair. Three cases with different coupling limit are considered:(a) Fixed Born-Infeld BHs endowed with charged scalar clouds, in which the scalar field is decoupled from both the metric and electromagnetic field; (b) Fixed Schwarzschild BHs with charged scalar clouds, in which the scalar field is coupled to the electromagnetic field but both are decoupled from the metric; (c) Static spherical charged BHs with charged scalar hair, in which all three fields are coupled. For each case, series of solutions are constructed numerically and their physical quantities (ADM mass $M$, electric charge $Q_e$ and hairiness $h$) are computed. Our main focus is on the effects of the non-linearity of the electromagnetic field on the solutions. Results show that increasing the Born-Infeld coupling constant $b$ will make the domain of existence of the solution shrink or even disappear when $b$ is large enough. This implies that, competing with the gravitation, nonlinearity of the electromagnetic field will make the formation of scalar cloud/hair harder or even impossible. Moreover, $b$ has significant influences on the profiles of the field configurations and physical quantities of the solutions. We should emphasis that the cloud/hair considered here is secondary which is very different from the scalar condensation in holographic superconductor model in asymptotically AdS space. For the later case, the scalar hair results from near-horizon tachyonic instability and is primary \cite{Hartnoll:2008vx,Sheykhi:2015mxb,Dias:2016pma,Rahmani:2020lju}.

 In this work, the scalar field we considered is complex and minimally coupled to the electromagnetic field. In past decades, there are also amounts of work on searching charged black hole solutions with a real scalar (or dilaton) hair within Einstein-Born-Infeld gravity (see for example Refs. \cite{Clement:2000ue,Tamaki:2001vv,Yazadjiev:2005za,Stefanov:2007qw,Sheykhi:2007gw,Wang:2020ohb}  and references therein). In these work, for the existence of hair, the real scalar (or dilaton) field is required to be non-minimally coupled to the electromagnetic field. Moreover, most of reported hairy solutions are not asymptotically flat with a few exceptions (see for example Refs. \cite{Stefanov:2007qw,Wang:2020ohb}).

There remains several interesting questions and possible further extensions left for future investigations. The first one is whether these hairy BHs are stable dynamically or thermodynamically. This requires careful studies of the classical stability of these BHs under perturbations and also of their phase structures. The second one is to extend the current research to other situations, such as other types of scalar potentials, higher-dimensional cases, other types of cloud/hair etc, to see the effects of non-linearity of the electromagnetic field.

	\begin{acknowledgments}
		
		The author would like to thank the anonymous referee for the valuable comments which helped to improve the manuscript. This work is supported by the National Natural Science Foundation of China (NNSFC) under Grant no.12075207.
		 
	\end{acknowledgments}

\end{document}